\documentclass[11pt]{article}
\usepackage[latin1]{inputenc}
\usepackage[english]{babel}
\usepackage[namelimits]{amsmath}
\usepackage{amssymb}
\usepackage{amsmath}
\usepackage{amsthm}

\begin{document}
\title{Axial and polar gravitational wave equations in a de Sitter expanding universe by Laplace transform.}
\author{
Stefano Viaggiu,\\
Dipartimento di Matematica,
Universit\`a di Roma ``Tor Vergata'',\\
Via della Ricerca Scientifica, 1, I-00133 Roma, Italy.\\
E-mail: {\tt viaggiu@axp.mat.uniroma2.it}}
\date{\today}\maketitle

\begin{abstract}
In this paper we study the propagation in a de Sitter universe of gravitational waves
generated by
perturbating some unspecified spherical astrophysical object in the frequencies domain.
We obtain the axial and polar perturbation equations in a cosmological de Sitter universe in the usual comoving 
coordinates, the coordinates we occupy in our galaxy. 
We write down the relevant equations in terms of Laplace transform with respect to the comoving time $t$
instead of the usual Fourier one that is no longer
available in a cosmological context. Both axial and polar perturbation equations are expressed in terms of a 
non trivial mixture of
retarded-advanced metric coefficients with respect to the Laplace parameter $s$ (complex translation). 
The axial case is studied in more detail.
In particular, the axial perturbations
can be reduced to a master linear second-order differential equation in terms of the Regge-Wheeler function $Z$ where a coupling with
a retarded $Z$ with respect to the cosmological time $t$ is present. It is shown that a de Sitter expanding universe can change the 
frequency $\omega$ of a gravitational wave as perceived by a comoving observer. 
The polar equations are much more involved. 
Nevertheless, we show that also the polar perturbations can be expressed in terms of four independent integrable differential equations.
\end{abstract}

\section{Introduction}
The recent detection of gravitational waves \cite{1} (GW150914 event) represents the birth of the gravitational wave astronomy.
The usual mathematical treatment for the gravitational wave equations (see for example \cite{1}-\cite{10} and references therein) assumes 
an asymptotically flat time-independent background. Einstein's equations are thus linearized about the chosen background. 
However, the standard cosmological model is formulated in terms of a Friedmann flat metric 
equipped with a positive cosmological constant $\Lambda$. Gravitational wave perturbation equations have been obtained and solved in \cite{11}
by using Fourier transform in the static patch of a de Sitter universe. Their equations show that effects due to $\Lambda$ on these equations are negligible. 
However, it is natural to ask about potential effects arising in a non-static universe. In this case, the background is no 
longer time independent
and the usual approach using the Fourier transform is generally not applicable since the metric coefficients are no longer
Fourier integrable with respect to $t$. In particular, in a de Sitter universe, the presence of the unperturbed scale factor
$a(t)=e^{Ht}, H^2=\Lambda/3$ makes the use of the Fourier transform problematic.\\
For an application of gravitational perturbations in a cosmological context 
relevant to the quantum primordial spectrum the reader should refer to the seminal papers by Grishchuk (see for example \cite{11a,11b}).

The issue concerning the formulation of a suitable mathematical framework for a gravitational wave theory in the non-static
patch of a de Sitter spacetime has been addressed in the papers \cite{12,13,14,15}. There, the authors 
analyze in great detail the asymptotic structure leading to a suitable mathematically sound formulation allowing a study of
gravitational waves propagating in a de Sitter universe.
Their equations substantially
confirm that also in the non-static patch of a de Sitter universe observable effects caused by the present-day value of 
$\Lambda$ are negligible. 
Recently \cite{16}, the attention focused on possible effects caused by 
$\Lambda$ 
on gravitational waves traveling in a de Sitter universe. The author of \cite{16}, 
for the purpose of write down and integrate the perturbation equations
Fourier transformed with respect to the time coordinate,
considered a de Sitter spacetime 
expressed in the Bondi-Sachs formalism. 
Since in the Bondi-Sachs formalism the metric is independent from the time coordinate $u$, he can solve the relevant equations
using Fourier transform. Also in this case, the conclusion is that the effects caused by $\Lambda$ are small provided that the frequency of the traveling wave is not too low. 
In the de Sitter background, the Bondi-Sachs coordinates can be easily obtained and so also the perturbated form of the metric.
As noticed in \cite{16}, to express a generic metric in the
Bondi-Sachs form a numerical integration is requested,
making this method no longer suitable for cosmological backgrounds
different from the de Sitter one.

From the reasonings above, the necessity
is thus evident of a deep understanding of the features of gravitational waves in a cosmological context. 
The usual approach \cite{4}-\cite{11} in terms of tensorial spherical harmonics has proved very succesful to study perturbations of
static and stationary spherical stars and black holes and to study the spectrum of frequencies of quasi-normal modes in terms of the famous Regge-Wheeler and Zerilli master equations. In a cosmological context \cite{16a,16b}, 
in order to study 
the spectrum of primordial inflation,
the interest mainly focused on cosmological perturbations. In this approach, with the spatial translational invariance of the background metric,
a Fourier transform in the wavenumber space ${\bf k}$ is performed and the perturbation metric functions
$h_{ik}(r,t)$ can be studied in the space $(k,t)$. However, it is also interesting both theorically and practically 
to wonder how a cosmological background can affect the propagation of a gravitational wave in the 
frequencies space $\omega$ (conjugate to $t$)
rather than in the wavenumber space ${\bf k}$, by mimicking the usual appraches of the stationary cases \cite{4}-\cite{11}.
This can be useful to study, for example, the quasi-normal modes emitted by a spherical star or by a black hole embedded in an expanding
universe. This paper is the first attempt to generalize the formalism used in the stationary case for the
search of quasi-normal modes derived in the seminal works \cite{3,4,5} to a cosmological time dependent context.
Moreover, our attention is not focused on the study of cosmological perturbations and the related topic 
concerning the evolution 
of cosmic structure or primordial fluctuations during inflation, although linear tensorial modes (i.e. gravitational waves)
are generated during the primordial inflation and some relation between the results of this paper and the usual ones can be
outlined.

Unfortunately, a cosmological background (Friedmann universes) is time dependent and the usual approach used for the 
stationary case in the frequencies space by means of Fourier transform is rather problematic. Our mainly purpose is thus to study a
possible generalization of the usual technique in the frequencies space. We show that this is possible by using,
from the onset, the Laplace transform.
 
Since papers \cite{12,13,14,15,16} analyse the physical effects caused by $\Lambda$ in a de Sitter universe, our attention is mainly focused 
on the mathematical derivation and study of the relevant equations. The study of propagation of gravitational waves in an expanding universe
can be found in \cite{a,b}. In \cite{a} the authors write down the axial equations in a generic Friedmann universe in relation to the 
Huygens principle. In \cite{b} also the polar case is considered for a de Sitter universe. In both cases the authors use conformal time and manage the equations by using the Fourier transform in wavenumber space conjugate to the spatial coordinate $r$
by Fourier transform.

Our main goal in this paper is to study the perturbation equations in the frequencies space for a 
gravitational wave traveling in a de Sitter universe and generated from some unspecified 
astrophysical source (binary system, black hole merging..). These perturbations are done with respect to the comoving cosmic time
$t$. This choice is due to the fact that we want to describe possible effects on the frequencies domain on a traveling 
wave caused by the cosmological constant as perceived by our point of observation within a comoving galaxy with our proper time
$t$.

In section 2 we specify the gauge, while in section 3 the linearization procedure of the relevant equations is summarized in detail.
In section 4 we write down the axial equations and in section 5 we study the effect on the frequencies space due to the cosmological constant.
In section 6 a study of the solution at the order 
$o(H)$ is done.
In section 7 we briefly analyse the integrability structure of the polar equations. Finally, section 8 is devoted to some conclusions and final remarks.

\section{The perturbed metric}

The starting point of our study is the de Sitter metric expressed in the usual comoving coordinates (we take $G=c=1$)
\begin{equation}
ds^2=dt^2-e^{2Ht}\left(r^2\sin^2\theta\;d{\phi}^2+dr^2+r^2 d{\theta}^2\right).
\label{1}
\end{equation}
We indicate with $g_{ik}^{(0)}$ the unperturbed metric (\ref{1}) and with 
$h_{ik}$ a small perturbation $|h_{ik}|<<|g_{ik}^{(0)}|$:
\begin{equation}
g_{ik}=g_{ik}^{(0)}+h_{ik}.
\label{2}
\end{equation} 
As usual, the spherical simmetry of $g_{ik}^{(0)}$ allows to write $h_{ik}$ in a basis of spherical 
tensorial harmonics \cite{4,5} expressed in terms of 
the spherical Legendre polynomials 
$Y_{\ell m}(\theta,\phi), \ell\in\mathbb{N}, m\in\mathbb{Z}, m\in [-\ell, +\ell]$. The polar perturbations 
$h_{ik}^{p}$  
under parity operator look like ${(-1)}^{\ell}$, while the axial ones
$h_{ik}^{a}$ look like ${(-1)}^{\ell+1}$.
We can specify four gauge conditions to simplify the expressions for 
$h_{ik}=h_{ik}^p+h_{ik}^a$. We use the diagonal gauge present in \cite{6,7,8,9,10}.
Hence, for $h_{ik}^a$ we have
\begin{equation}
h_{ik}^a=
\begin{pmatrix}
(t) & (\phi) & (r) & (\theta)\\
0 & h_0\sin\theta\;Y_{\ell m,\theta} & 0 & -h_0\frac{1}{\sin\theta}Y_{\ell m,\phi}\\
h_0\sin\theta\;Y_{\ell m,\theta} & 0 & h_1\sin\theta\;Y_{\ell m,\theta} & 
0\\
0 & h_1\sin\theta\;Y_{\ell m,\theta} & 0 & -h_1\frac{1}{\sin\theta}\;Y_{\ell m,\phi}\\
-h_0\frac{1}{\sin\theta}Y_{\ell m,\phi} & 0 & -h_1\frac{1}{\sin\theta}\;Y_{\ell m,\phi} &
0
\end{pmatrix},
\label{3}
\end{equation}
while for $h_{ik}^p$ we obtain
\begin{equation}
h_{ik}^p=
\begin{pmatrix}
(t) & (\phi) & (r) & (\theta)\\
2N Y_{\ell m} & 0 & 0 & 0\\
0 & -2r^2\sin^2\theta e^{2Ht}\;H_{11} & 0 & -r^2 V X_{\ell m}\\
0 & 0 & -2 e^{2Ht}L Y_{\ell m} & 0\\
0 & -r^2 V X_{\ell m} & 0 & -2 r^2 e^{2Ht} H_{33}
\end{pmatrix}. 
\label{4}
\end{equation}
The axial perturbations (\ref{3}) depend on two functions
$h_0(t,r),h_1(t,r)$, 
while the polar perturbations (\ref{4}) depend on four functions $N(t,r),L(t,r),T(t,r),V(t,r)$. Moreover
\begin{eqnarray}
X_{\ell m}(\theta,\phi) &=& 2Y_{\ell m,\theta,\phi}-2\cot\theta\;Y_{\ell m,\phi}\label{5}\\
W_{\ell m}(\theta,\phi) &=& Y_{\ell m,\theta,\theta}-\cot\theta\;Y_{\ell m,\theta}-
\frac{1}{\sin^2\theta}\;Y_{\ell m,\phi,\phi}\nonumber\\
H_{11}(t,r,\theta,\phi) &=&TY_{\ell m}+\frac{V}{\sin^2\theta}\;Y_{\ell m,\phi,\phi}+
V\cot\theta\;Y_{\ell m,\theta}\nonumber\\
H_{33}(t,r,\theta,\phi) &=&TY_{\ell m}+VY_{\ell m,\theta,\theta}.
\end{eqnarray} 
The line element thus becomes
\begin{eqnarray}
ds^2 &=& dt^2-e^{2Ht}\left(r^2\sin^2\theta\;d{\phi}^2+dr^2+r^2 d{\theta}^2\right)+\label{6}\\
&+&\sum_{\ell m}\Bigl\{2N(t, r) Y_{\ell m}dt^2-
2e^{2Ht}L(t,r) Y_{\ell m}dr^2-\nonumber\\
&-& 2r^2\sin^2\theta e^{2Ht}H_{11}(t, r,\theta,\phi)d{\phi}^2-2r^2 e^{2Ht}H_{33}(t, r,\theta,\phi)d{\theta}^2+\nonumber\\
&+& 2h_0(t,r)\sin\theta\;Y_{\ell m,\theta}\;dt d\phi-
2h_0(t,r)\frac{Y_{\ell m,\phi}}{\sin\theta}dt d\theta+\nonumber\\
&+& 2h_1(t,r)\sin\theta\;Y_{\ell m,\theta}\;dr d\phi-
2h_1(t,r)\frac{Y_{\ell m,\phi}}{\sin\theta}dr d\theta-\nonumber\\
&-& \left[4r^2 V(t,r) Y_{\ell m,\theta,\phi}-4r^2 V(t,r)\cot\theta\;Y_{\ell m,\phi}\right]d\theta d\phi\Bigr\}.\nonumber
\end{eqnarray}
In the following, we use the metric (\ref{6}) to obtain linearized Einstein's equations for the perturbed metric coefficients.

\section{Perturbation of the field equations and Laplace transform}

A convenient way to simplify the perturbed equations is to introduce a tetradic basis of four vectors $e_{(a)i}$ ($a=t,r,\theta,\phi $),
with the usual property that  $e_{(a)}^i e_{(b)i}={\eta}_{(a)(b)}$ with ${\eta}_{(a)(b)}=diag(1,-1,-1,-1)$.
The field equations are thus proiected onto $e_{(a)i}$, i.e. $G_{(a)(b)}=2T_{(a)(b)}+\Lambda{\eta}_{(a)(b)}$, and perturbed. We have
$\delta G_{(a)(b)}=2\delta T_{(a)(b)}+\delta(\Lambda{\eta}_{(a)(b)})$. For the unperturbed metric we have
\begin{eqnarray}
& & e_{t}^i=(1,0,0,0), \label{7}\\
& & e_{\phi}^i=\left(0,\frac{1}{re^{Ht}\sin\theta},0,0\right), \nonumber\\
& & e_{r}^i=\left(0,0,\frac{1}{e^{Ht}},0\right), \nonumber\\
& & e_{\theta}^i=\left(0,0,0,\frac{1}{r e^{Ht}}\right), \nonumber\\
& & G_{tt}=3H^2, G_{\phi\phi}=-3H^2 r^2\sin^2\theta e^{2Ht}, G_{rr}=-3H^2 e^{2Ht}, G_{\theta\theta}=-3r^2H^2 e^{2Ht}.\nonumber
\end{eqnarray}
For the perturbed metric we have \footnote{In all cases we consider only perturbations associated to gravitational waves, i.e. $\ell \geq 2$.}:
\begin{eqnarray}
& &\delta G_{(a)(b)}=\delta\left(e_{(a)}^i e_{(b)}^k G_{ik}\right)=2\delta\left(e_{(a)}^i e_{(b)}^k T_{ik}\right),\label{8}\\
& &\delta G_{ik}=\delta R_{ik}-\frac{h_{ik}}{2}R+\frac{1}{2}g_{ik}^{(0)}R_{lm}h^{lm}-
\frac{1}{2}g_{ik}^{(0)}g^{(0)lm}\delta R_{lm}\nonumber,
\end{eqnarray}
and also
\begin{eqnarray}
& & \delta e_{(t)}^i=\left[-N Y_{\ell m},\frac{h_0}{r^2e^{2Ht}\sin\theta}Y_{\ell m,\theta},0,-\frac{h_0}{r^2e^{2Ht}\sin\theta}
Y_{\ell m,\phi}\right],\nonumber\\
& & \delta e_{(\phi)}^i=\left[0,-\frac{H_{11}}{re^{Ht}\sin\theta},0,0\right],\nonumber\\
& & \delta e_{(r)}^i=\left[0,\frac{h_1}{r^2e^{3Ht}\sin\theta}Y_{\ell m,\theta},-\frac{L}{e^{Ht}}Y_{\ell m},
-\frac{h_1}{r^2e^{3Ht}\sin\theta}Y_{\ell m,\phi}\right],\nonumber\\
& & \delta e_{(\theta)}^i=\left[0,-\frac{V}{re^{3Ht}\sin^2\theta}X_{\ell m},0,-\frac{H_{33}}{re^{Ht}}\right]. \label{9}
\end{eqnarray}
Concerning the energy momentum tensor $T_{ik}$, for a de Sitter universe we obviously have $\delta T_{ik}=0$. 
We are considering gravitational waves traveling in a de Sitter universe.
Hence, the generation of a gravitational wave is not a consequence of the perturbation of $\Lambda$, but the gravitational wave 
perturbates the spacetime metric\footnote{In practice, the gravitational wave is assumed to have no interaction
	with $\Lambda$.}. We thus have
\begin{equation}
\delta G_{(a)(b)}= -\delta(\Lambda\;{\eta}_{(a)(b)})=-\Lambda\;\delta{\eta}_{(a)(b)}=0.
\label{10}
\end{equation}
In (\ref{10}) we used the fact that ${\eta}_{(a)(b)}$ is a constant metric tensor\footnote{This is the reason for which 
perturbation equations simplify in a tetradic formalism.}. Moreover, note that if we consider a Friedmann metric equipped with a 
perfect fluid $T_{\mu\nu}=(\rho+p)u_{\mu}u_{\nu}-pg_{\mu\nu}$, the term $\Lambda g_{\mu\nu}$ can be obviously seen as a perfect fluid with
$\rho=-p,\{\rho,p\}\in\Re$. However, in the case of a gravitational wave traveling in a Friedmann universe with
$T_{\mu\nu}\neq 0$, we generally expect that $\{\delta\rho,\delta p\}\neq 0$ since the perturbed hydrostatic equation \cite{6,7}
implies that $\delta p\sim(\rho+p)$ (that is vanishing for the de Sitter case).

In the usual approach, the metric coefficients $h_{ik}$ are Fourier transformed with respect to the frequency $\omega$: the real part 
$\omega$ denotes the proper frequency of the wave, while the complex part the damping rate. Unfortunately, for the metric (\ref{1}) and more generally in a time dependent cosmological context, Fourier transform cannot be available. In fact, the expansion factor 
$a^2(t)=e^{2Ht}$ is no longer Fourier integrable, in particular for
$t\in[-\infty,\infty]$. However, it is Laplace transformable, provided that the 
abscissa of convergence $a_0$ is greater than $2H$ with $Le^{2Ht}(s)=\int_{0}^{\infty} e^{2Ht} e^{-st}dt$.
The more convenient strategy for the metric (\ref{1}) is the following. First of all, we must obtain the perturbation equations by
taking the time dependence left explicit. Only after the equations have been obtained, we use Laplace antitransform. As an example, by 
denoting with $A(t,r)=\{N,L,T,\cdots\}$ a generic perturbed metric coefficient of (\ref{3}) or (\ref{4}), the following texture will be used:
\begin{eqnarray}
& & e^{\pm 2Ht} A(t,r)=e^{\pm 2Ht}\frac{1}{2\pi\imath}\lim_{b\rightarrow +\infty}\int_{a_0-\imath b}^{a_0+\imath b}A(s,r)e^{st}ds=\nonumber\\
& & =\frac{1}{2\pi\imath}\lim_{b\rightarrow +\infty}\int_{a_0-\imath b}^{a_0+\imath b}A(s\mp 2H,r)e^{st}ds,\label{11}\\
& & e^{\pm 2Ht} A_{,t,t}(t,r)=
\frac{1}{2\pi\imath}\lim_{b\rightarrow +\infty}\int_{a_0-\imath b}^{a_0+\imath b}{(s\mp 2H)}^2 A(s\mp 2H,r)e^{st}ds,\nonumber
\end{eqnarray}
provided that $\Re(s) > 2H, a_0>2H$ and after using known properties of Laplace transform.  
Since by definition $|\{A(t,r)\}|<<1$,
the perturbed metric
coefficients $\{A(t,r)\}$ are supposed to be Laplace transformable 
(together with their partial derivatives)
$\forall s>0$. Laplace transformability in presence of the expansion factor $e^{2Ht}$ implies that $\Re(s)> 2H$.

Moreover, the $\{A(t,r)\}$ are supposed to have non-vanishing support for a given initial time $t_i$ that can be chosen,
without loss of generality, to be zero.
This is a realistic situation because perturbations are generated at a given finite time. However, since we work with Laplace transformate
$\{A(s,r)\}$, the initial value plays no role.

The advantages of the use of Laplace transform with respect to other approaches are:
\begin{enumerate}
\item We can adopt the usual comoving coordinates where Friedmann metrics assume simple expressions.
\item Since a large class of cosmological expansion factors $a(t)$ are Laplace transformable, there is hope to generalize our method
to a generic cosmological model.
\item In a cosmological context the universe begins with a big bang at $t=0$ or at $t=0$ with a de Sitter phase. Hence the perturbations
cannot be integrated in the whole range $t\in(-\infty, +\infty)$
\item The perturbating equations are expressed in a clear gauge and as a result the identification of tensorial modes with
respect to gauge modes can be made without any ambiguity. 	
\end{enumerate}	
To the best of our knowledge, the use of the Laplace transform in a cosmological context is completely new.

In the next sections, we apply the technique depicted above. We analyze 
in great detail the axial case, since the linearized axial equations are more simple to manage
with respect to the polar ones. Nevertheless,
also for the polar case we show that the system of equations is integrable. We stress that our attention is 
mainly focused on the method rather than on the effects of $\Lambda$ on the traveling waves: for this purpose the reader can see the 
very interesting papers 
\cite{12,13,14,15,16}. Concerning the notation, in the following sections, with $h_0(t,r),h_1(t,r)\cdots$ we mean
$h_{0\ell m}(t,r),h_{1\ell m}(t,r)\cdots$: in practice the dependence of the perturbed metric coefficients from the Legendre indices
$\ell,m$ is implied.

\section {Axial perturbations}

The relevant equations for the polar perturbations are given by
the tetradic components $(t)(\phi), (\phi)(r),(\phi)(\theta)$. After the usual variables separation we obtain:
\begin{eqnarray}
& & -\frac{(n+1)h_0(t,r)}{r^2}+\frac{h_{0,r,r}(t,r)}{2}+H h_{1,r}(t,r)+\nonumber\\
& &+\frac{2H h_1(t,r)}{r}-\frac{h_{1,t}(t,r)}{r}-\frac{h_{1,t,r}(t,r)}{2}=0,\label{12}\\
& & -\frac{h_{0,t}(t,r)}{r}+\frac{h_{0,t,r}(t,r)}{2}-\frac{H h_{0,t}(t,r)}{r}+\frac{H h_{0,r}(t,r)}{2}
-\frac{n h_{1}(t,r)}{r^2}e^{-2Ht}-\nonumber\\
& &-\frac{h_{1,t,t}(t,r)}{2}+H^2h_1(t,r)+\frac{H h_{1,t}(t,r)}{2}=0,\label{13}\\
& & h_{1,r}=H h_0 e^{2Ht}+ e^{2Ht} h_{0,t}, \label{14}
\end{eqnarray}
where $2n=(\ell-1)(\ell+2)$. For $H=0$ the equations (\ref{12})-(\ref{14}) reduce to the ones in \cite{8}. After 
using Laplace antitransform and following the notation of equations (\ref{11}) we have:
\begin{eqnarray}
& & -\frac{(n+1)}{r^2}h_{0}(s,r)+\frac{h_{0,r,r}(s,r)}{2}+H h_{1,r}(s,r)+\frac{2H}{r}h_1(s,r)-\nonumber\\
& & -\frac{s}{r}h_1(s,r)-\frac{s}{2}h_{1,r}(s,r)=0,\label{15}\\
& & -\frac{s}{r}h_0(s,r)+\frac{s}{2}h_{0,r}(s,r)-\frac{H}{r}h_0(s,r)+\frac{H}{2}h_{0,r}(s,r)-\frac{n}{r^2}h_1(s+2H,r)-\nonumber\\
& & -\frac{s^2}{2}h_1(s,r)+H^2 h_1(s,r)+\frac{sH}{2}h_1(s,r)=0,\label{16}\\
& & h_{1,r}(s+2H,r)=H h_0(s,r)+s h_0(s,r). \label{17}
\end{eqnarray}
Equation (\ref{17}) can be written in the equivalent form:
\begin{equation}
h_{1,r}(s,r)=(s-H) h_0(s-2H,r).
\label{18}
\end{equation}
In the usual treatment using Fourier transform in a stationary asymptotically flat spacetime, 
equations (\ref{15}) and (\ref{16})
are dependent and the independent equations chosen are the (\ref{16}) and the (\ref{17}). In our case, we have three equations depending on 'retarded'
and 'advanced' functions with respect to the Laplace parameter $s$\footnote{Since this behaviour does not happen in
the time domain $t$, this 'retarded' and 'advanced' dependence of the field equations  merely does imply translations 
with respect to the complex parameter $s$.}. To start with,
consider equation (\ref{16}) derived with respect to ''$r$'' and subtract the equation so obtained to the (\ref{15})
multiplied by $(s+H)$. As a result we obtain, after using
the (\ref{17}), the trivial identity $0=0$. Hence, also in presence of $H$, equations (\ref{15}) and (\ref{16}) are dependent. 
Hence, we have at our disposal equations (\ref{16}) and (\ref{17}).  
We are now in the position to write down the generalized Regge-Wheeler equation present in 
\cite{8}. We can use the (\ref{17}) to express $h_0(r)$ in terms
of $h_{1,r}(s+2H,r)$
After introducing the Regge-Wheeler function $Z(s,r)$ as $h_1(s,r)=rZ(s,r)$ and its retarded counterpart as
$h_1(s+2H,r)=rZ(s+2H,r)$, we obtain 
\begin{equation}
Z_{,r,r}(s+2H,r)-2\frac{(n+1)}{r^2}Z(s+2H,r)=
(s+H)(s-2H)Z(s,r),{\label{19}}
\end{equation}
or 
\begin{equation}
Z_{,r,r}(s,r)-2\frac{(n+1)}{r^2}Z(s,r)=
(s-H)(s-4H)Z(s-2H,r),{\label{20}}
\end{equation}
Equations (\ref{19}) and (\ref{20}) obviously reduce for $H=0$ to the usual Regge-Wheeler equation in a Minkowski spacetime
\cite{8}.

\section{A frequency study}

The use of Laplace transform to study gravitational waves is the major novelty present in this paper. 
In this section we show that this new approach allows one to obtain some interesting results concerning the
perceived frequency of a traveling gravitational wave.  
As stressed at the introduction, the use of Laplace transform is essential in order to study the effects due to the expanding
universe on the frequencies spectrum of a traveling gravitational wave generated from 
some unspecified astrophysical source, outside the astrophysical source. The perturbations are expanded in the frequencies space
with respect to the cosmological proper time $t$, the time we measure within our galaxy: this is the natural choice if we 
consider a frequency study.\\
For the actual small value of the cosmological constant, as shown in \cite{16}, these corrections give small effects. However, the
coupling between $Z(s,r)$ and $Z(s-2H,r)$ takes evident, as shown in \cite{12,13,14,15}, 
the difficulty of identifying a gravitational wave in a cosmological context. In particular, the presence of 
$H$ changes the usual expression of the wave equation suitable for
a wave propagating in a flat Minkowski spacetime that in Laplace transform is given by
\begin{equation}
Z_{,r,r}(s,r)-\frac{\ell(\ell+1)}{r^2}Z(s,r)-s^2Z(s,r)=0.
\label{v1}
\end{equation}
With respect to the (\ref{v1}), consider a gravitational wave propagating in the vacuum with frequency 
$\omega\in\Re$.\\ 
How we can physically consider the parameter $s$ ? 
First of all, we can obviously  write
$s=-\imath\omega+s_0$, where $|\omega|$ can be seen as the real (angular ) frequency of propagation of the Fourier mode $\omega$.
In a Minkowski spacetime $Z(r,t)$ is Fourier integrable and by setting 
$s_0=0$ in (\ref{v1}) we regain the expression present in \cite{8}. In fact, for a function whose abscissa of convergence is 
zero the (bilateral) Laplace transform coincides with the Fourier one. Thanks to the meaning of the abscissa of
convergence $s_0$, the Laplace antitransform converges, apart from a zero measure Lebesgue set, to the original function 
$\forall \Re(s) \geq s_0$. What seems physically relevent is the $Inf\{\Re(s)\}$. In fact, for the (\ref{v1}) 
the physical frequency related to the mode $s$ is obtained by considering $s_0=0, \omega\in\Re$. Hence,
the real angular frequency, perceived (measured) by a physical observer at rest in a Minkowski vacuum,
is given by the following formula:
\begin{equation}
{\omega}^2=\lim_{\Re(s)\rightarrow 0}(-s^2).
\label{v2}
\end{equation}
In our context, since of the presence of $e^{2Ht}$ in the metric and in (\ref{20}), we have
$\Re(s)>2H$. 
According to the reasonings above, we may assign to $s_0$, and in particular to the $Inf\{\Re(s)\}$, a physical 
meaning. In practice, a non-vanishing $s_0$ means that the frequency really measured by an observer (comoving) with respect
to the time $t$ is no longer given by $\omega$, but it is modified by an amount depending on $Inf\{\Re(s)\}$. 
This implies that the expansion of the universe can
modify the frequency of emission $\omega$, as measured by a comoving observer,
by a quantity that, according to the equation (\ref{20}) with the term $Z(s-2H,r)$,
it depends on the $Inf_{\Re(s)}$. If this interpretation is correct,
we can thus define an effective perceived angular frequency ${\omega}_p$ given by
\begin{equation}
{\omega}_p^2=\Re\left(\lim_{\Re(s)\rightarrow 2H}-s^2\right)=\omega^2-\frac{4}{3}\Lambda.
\label{v3}
\end{equation}
The formula (\ref{v3}) shows that the modes generated with angular frequency $\omega$ of the order of $\sqrt{\Lambda}$ are strongly modified by the expansion 
of the universe. Since $\sqrt{\Lambda}\sim 10^{-18}Hz$, we expect 
that ordinary sources are not affected by the expansion of the universe. We also expect that modes with $\omega\sim H$ have practically ${\omega}_p\sim 0$, i.e. gravitational waves
propagating with frequency $\omega\sim H$ are perceived as frozen (standing waves) by a comoving observer.
Conversely, modes with $\omega>>H$, as suggested by 
physical intuition, are practically left unchanged by a de Sitter expanding universe. Moreover, if $s_0<0$ then the perturbation
is Fourier integrable and and it is equivalent to the case $Inf_{\Re(s)}=0$. 
These considerations in the frequency domain can be compared with the usual ones obtained in the wavelength domain by considering
tensorial perturbations (i.e. gravitational waves) during primordial inflation
\cite{16a,16b,17}.
In the usual approach concerning tensorial perturbations of a de Sitter spacetime or of 
a generic Friedmann flat spacetime, the following form of the metric is used in terms of the conformal time
$\eta$:
\begin{equation}
ds^2=a^2(\eta)\left[{d{\eta}}^2-({\delta}_{ij}+h_{ij})dx^i dx^j\right],
\label{v4}
\end{equation}
where generally the perturbating term $h_{ij}$ is a traceless tensor described by two degrees of polarization
\footnote{Note the analogy with the axial case described by the transverse gauge given by (\ref{3}).} $h_1,h_2$.
The usual analysis \cite{16a,16b,17} can be obtained by Fourier expanding $h_1,h_2$ in the wavelength space $k$. As well known,
standard results imply that proper modes $k/a(\eta)$ such that $k/a(\eta)<<H/c$ are frozen.
Since for the proper wavelength ${\lambda}_p$ we have ${\lambda}_p\sim a(\eta)/k$, with
$k/a(\eta)<<H/c$ we have over (Hubble) horizon modes. Conversely, modes with  $k/a(\eta)>>H/c$ (under Hubble horizon modes)
are practically left unchanged by the de Sitter expansion.\\
As a consequence of these reasonings, our proposed interpretation of the Laplace parameter $s$, and in particular of its 
abscissa of convergence $s_0$, is in agreement with standard results obtained in the cosmological perturbations theory 
during primordial inflation. Similar results can be found in \cite{16} but with respect to a different time coordinate.
The advantage of our approach is thus to read the effects of $\Lambda$ on the frequencies measured with
respect to the proper cosmic time $t$ in a clear physical way.

If our physical interpretation of $s_0$ is correct, then some arguments on the nature of the cosmological constant can be
outlined. In fact, in \cite{vs} it has been proposed that in a de Sitter expanding universe the cosmological constant
$\Lambda$ can be seen as composed of gravitons near a Bose-Einstein condensate phase. The fact that 
modes with frequencies $\omega\sim H$ are frozen may be a consequence of the fact that in a de Sitter universe, gravitons
that eventually compose $\Lambda$, all have a mean frequency $\overline{\omega}$ of the order of 
$H\sim \sqrt{\Lambda}$ (see \cite{vs}). Hence, for a comoving observer gravitational
modes with $\omega\sim H$ could be indistinguishable from the background gravitons bath
and thus he/her observes a practically frozen wave.
This phenomenon may be the analogous of the one concerning light propagating in a medium. In particular, as well known,
a very cold medium can slow down an electromagnetic wave.\\ 
In a cosmological context, with a de
Sitter universe (cosmological constant) \cite{vs2} can be associated a Hawking temperature 
$T_h$ proportional to $\sqrt{\Lambda}$. With the actual very small value of $\Lambda$, we expect a very cold universe
with $T_h\sim 0$. Hence a gravitational wave with $\omega\sim 1/\sqrt{\Lambda}$ could be frozen by the very cold 
background.

At this point, it is thus natural to ask what could happen if the same study in frequencies domain is performed in a background 
different from the de Sitter one. According to the reasonings above, if we consider a background such that 
for $t\rightarrow\infty$ we have, for example, $a(t)>Me^{s_0t}$ with $M\in\Re, \forall s_0\in\Re$, perturbations
are no longer Laplace integrable and as a consequence a comoving observer should see a rapidely 
decaying wave.\\
Conversely, consider a Friedmann background with, for example, a power law expansion scale factor $a(t)$. In this case we
have $s_0=0$ and according to our interpretation for $s_0$, formula (\ref{v2}) holds true
and as a result the frequency perceived by the comoving observer is left unchanged, while obviously the equation for $Z(s,r)$ is not expected to be the one given by the vacuum case
(\ref{v1}). In fact, by considering for example the radiation case with $a(t)\sim\sqrt{t}$, instead of the
(\ref{14}) we have
\begin{equation}
h_{1,r}(t,r)=h_0(t,r)+2t\;h_{0,t}(t,r).
\label{v5}
\end{equation}
Hence, after Laplace transform application, instead of the (\ref{17}) we obtain
\begin{equation}
h_{1,r}(s,r)=-h_{0}(s,r)-2s\;h_{0,s}(s,r).
\label{v6}
\end{equation}
In equation (\ref{v6}) the 'retarded' behaviour $Z(s-2H,r)$ disappeared and instead we have ordinary derivatives with respect to
the Laplace parameter $s$ (see equation (\ref{23})).

Summarizing, according to our proposed interpretation for $s_0$, the de Sitter case is the only one where the frequency of the traveling gravitational wave as perceived from a comoving observer is changed with respect to the 'emission' value $\omega$. 
In practice the frequency $\omega$ can be seen as the one measured by a freefalling observer. Such
modification of $\omega$ is relevant when $\omega\sim H$ and negligible when $\omega>>H$. For Friedmann spacetimes 
filled with ordinary matter-energy, this phenomenon does not happen. This motivate the hypothesis that the
cosmological constant can be composed of gravitons with mean frequency $\overline{\omega}$ of the order of
$H$, as conjectured in \cite{vs}.  

\section{A perturbative study of the Regge-Wheeler equation}

First of all, note that the effects of $H$  are present also in the term multiplied by $s^2$. This is 
an interesting fact since in the 
Laplace transformed wave equation the term with $s^2$ ($-s^2$ in Fourier transform with $\Re(s)=0$ )
is related to the term $Z_{,t,t}$. 
This means that the expansion acts in a non-trivial way also on the traveling gravitational wave.
Mathematically, this means that the modified wave equation can be written as:
\begin{equation}
\left[-e^{2Ht}{\partial}_{tt}+{\partial}_{rr}-\frac{\ell(\ell+1)}{r^2}\right]Z(t,r)+e^{2Ht}\left[H{\partial}_{t}+2H^2\right]Z(t,r)=0.
\label{20b}
\end{equation}
Equation (\ref{20b}) can also be written in the equivalent way:
\begin{equation}
\left[-{\partial}_{tt}+e^{-2Ht}{\partial}_{rr}-e^{-2Ht}\frac{\ell(\ell+1)}{r^2}\right]Z(t,r)+\left[H{\partial}_{t}+2H^2\right]Z(t,r)=0.
\label{21b}
\end{equation}
The (\ref{21b}) is a consequence of variables separation.\\ 
By considering $Z(t,r,\theta,\phi)=Z(t,r)C_{\ell+2}^{-\frac{3}{2}}(\theta)e^{\imath m\phi}$, where $C_{\ell+2}^{-\frac{3}{2}}(\theta)$
denote Gegenbauer functions \cite{3}, we have for the (\ref{21b})
\begin{eqnarray}
& & -{\Box} Z(t,r,\theta,\phi)+\left[HZ_{,t}(t,r,\theta,\phi)+2H^2 Z(t,r,\theta,\phi)\right]=0, \nonumber\\
& & {\Box}=g^{\mu\nu}{\nabla}_{\mu}{\nabla}_{\nu}=-e^{-2Ht}{\nabla}^2+\frac{{\partial}^2}{\partial t^2},\label{21bb}
\end{eqnarray}
where $\Box$ is the D'Alambertian operator in the de Sitter spacetime (\ref{1}). The equation (\ref{21bb}) is similar to the one of scalar perturbations on a de Sitter background with potential function $V(Z)\sim H^2 Z^2$.
To solve the (\ref{21b}) one could, as shown in \cite{17}, to use the following ansatz. By using Cartesian coordinates, we have:
\begin{equation}
\sum_{k=0}^{\infty}\left[A_k(x,y,z)e^{-kHt}+B_k(x,y,z)t e^{-kHt}\right],
\label{22bb}
\end{equation}
where the solution can be obtained for $A_k$ and $B_k$ after insertion in the (\ref{21bb}).\\
In this paper we are mainly interested to the generalization of the usual technique suitable in
a static background in the frequency space to cosmological spacetimes and solving the relevant equations with respect to $r$.\\ 
Equation (\ref{20}) (or its rescaled form (\ref{29})), can be written as
\begin{eqnarray}
& & Z_{,r,r}(s,r)-\frac{\ell(\ell+1)}{r^2}Z(s,r)-s^2Z(s,r)=\label{21}\\
& & =s^2\left[Z(s-2H,r)-Z(s,r)\right]-H(5s-4H)Z(s-2H,r).\nonumber
\end{eqnarray}
The first member of the (\ref{21}) is nothing else but the usual Regge-Wheeler equation
in a Minkowski spacetime. Physically, the right hand side of (\ref{21}) can be seen as a forcing term describing the interaction 
between the traveling wave and the cosmological constant. The presence of the retarded function $Z(s-2H,r)$ in the parameter space $s$
indicates, as shown in section 5, that the expanding universe changes the frequency of the wave as perceived by a comoving observer.

We can use the actual very low value
of $H\sim 10^{-26}/m^2$ to perform some simplification. To this purpose, consider the term proportional to $s^2$ on the right side of
(\ref{21}). We can use the following approximation:
\begin{eqnarray}
& &\left[Z(s-2H,r)-Z(s,r)\right]=\frac{\left[Z(s-2H,r)-Z(s,r)\right]}{-2H}\left(-2H\right)=\nonumber\\
& &=-2H Z_{,s}(s,r)+o(H).\label{22}
\end{eqnarray}
As a result, at the order $H$ and neglecting terms $o(H)$, equation (\ref{21}) becomes:
\begin{equation}
Z_{,r,r}(s,r)-\frac{\ell(\ell+1)}{r^2}Z(s,r)-s^2Z(s,r)=-2Hs^2 Z_{,s}(s,r)-5HsZ(s,r).
\label{23}
\end{equation}
We can study the general solution of the (\ref{23}) in this approximation. We can expand the solution $Z(s,r)$ in powers of
$H$. After introducing a small parameter $\epsilon$, we have:
\begin{equation}
Z(s,r)=Z^{(0)}(s,r)+\epsilon Z^{(1)}(s,r)+{\epsilon}^2 Z^{(2)}(s,r)+\cdots+{\epsilon}^n Z^{(n)}(s,r).
\label{24}
\end{equation}
We must insert the (\ref{24}) in the (\ref{23}). At the order $\epsilon$, we thus have:
\begin{equation}
Z_{,r,r}^{(1)}(s,r)-\frac{\ell(\ell+1)}{r^2}Z^{(1)}(s,r)-s^2Z^{(1)}(s,r)=-2Hs^2 Z_{,s}^{(0)}(s,r)-5HsZ^{(0)}(s,r),
\label{25}
\end{equation}
where $Z^{(0)}(s,r)$ is the solution of the Regge-Wheeler equation for $H=0$. We can write the 
'known' solution for $Z^{(0)}$. We have:
\begin{eqnarray}
& & Z^{(0)}(s,r)=e^{(-sr)}\left[A_0+\frac{A_1}{r}+\frac{A_2}{r^2}+\cdots+
\frac{A_k}{r^k}\right]+\nonumber\\
& & +B_0e^{(sr)}\left[A_0-\frac{A_1}{r}+\frac{A_2}{r^2}+\cdots+
\frac{A_k}{r^k}\right]\nonumber\\
& & A_k=\frac{A_{k-1}}{2ks}\left[\ell(\ell+1)-k(k-1)\right], \label{26}
\end{eqnarray} 
where $B_0$ is a constant.
The general solution for the (\ref{25}) is given by the general solution of the homogeneous equation summed to the particular solution
of the inhomogeneous one.\\
We are now in the position to obtain the solution of (\ref{25}). In the next subsections we consider the quadrupole 
($\ell=2$) and the sextupole ($\ell=3$) cases.

\subsection{$\ell=2$}

For $\ell=2$ we have
\begin{eqnarray}
& & Z^{(0)}_{2m}=A_0 e^{-sr}\left[1+\frac{3}{sr}+\frac{3}{s^2r^2}\right]+ \label{s1}\\
& &+B_0 A_0 e^{sr}\left[1-\frac{3}{sr}+\frac{3}{s^2r^2}\right].\nonumber 
\end{eqnarray}
After inserting the (\ref{s1}) we obtain for the particular solution $Z_p^{(1)}$:
\begin{eqnarray}
& & Z_p^{(1)}=e^{-sr}\left[Q_1 r^2+Q_2 r+Q_3\right]+\label{s2}\\
& &  +Be^{sr}\left[Q_1 r^2-Q_2 r+Q_3\right],\nonumber\\
& & Q_1=-\frac{A_0 Hs}{2},\;Q_2=\frac{A_0 H}{2},\;Q_3=\frac{A_0 H}{2s},\nonumber
\end{eqnarray}
with $B\in\Re$.
Hence for the general solution we have
\begin{eqnarray}
& & Z_{2m}(s,r)=A e^{-sr}\left[1+\frac{3}{sr}+\frac{3}{s^2r^2}\right]+\label{s3}\\
& & + B e^{sr}\left[1-\frac{3}{sr}+\frac{3}{s^2r^2}\right]+\nonumber\\ 
& &+A_0 e^{-sr}\left[-\frac{sH}{2}r^2+\frac{H}{2}r+\frac{H}{2s}\right]+\nonumber\\
& & + B_0 e^{sr}\left[-\frac{sH}{2}r^2-\frac{H}{2}r+\frac{H}{2s}\right].\nonumber 
\end{eqnarray}
where $A$ is a generic constant. From a first inspection of the (\ref{s3}), we see that corrections caused by $H$ do appear with the dominant one 
proportional to $r^2$. Certainly for $rH<<1$ these corrections are completely negligible. Conversely, for $rH\sim 1$, this term becomes no longer
negligible and effects on the propagation of the gravitational wave can arise, although a more rigorous coordinate independent study
\cite{16} shows that these effects can be ignored from the point of view of the detection of gravitational waves. Also note that a de Sitter 
universe is equipped with apparent horizon and future event horizon. There, as well known, the asymptotic behaviour is useless. At the actual 
cosmological time $t_0$, we have that the proper areal radius of the universe looks like $L\sim 1/H$ and in our case, since 
$e^{Ht}$ is of the order of unity, $r\sim 1/H$ is of the order of the proper radius of our visible universe. 

\subsection{$\ell=3$}

For $\ell=3$ we have
\begin{eqnarray}
& & Z^{(0)}_{3m}=A_0 e^{-sr}\left[1+\frac{6}{sr}+\frac{15}{s^2r^2}+\frac{15}{s^3 r^3}\right]+ \label{ss1}\\
& & + B_0 e^{sr}\left[1-\frac{6}{sr}+\frac{15}{s^2r^2}-\frac{15}{s^3 r^3}\right].\nonumber
\end{eqnarray}
With the same technique of the subsection above we obtain:
\begin{eqnarray}
& & Z_{3m}(s,r)=A e^{-sr}\left[1+\frac{6}{sr}+\frac{15}{s^2r^2}+\frac{15}{s^3 r^3}\right]+\label{ss2}\\
& & + B e^{sr}\left[1-\frac{6}{sr}+\frac{15}{s^2r^2}-\frac{15}{s^3 r^3}\right]-\nonumber\\
& & -A_0 e^{-sr}\left[\frac{sH}{2}r^2+H r+\frac{3H}{2s}+\frac{3H}{2s^2r}\right]+\nonumber\\
& & + B_0 e^{sr}\left[\frac{sH}{2}r^2-H r+\frac{3H}{2s}-\frac{3H}{2s^2r}\right].\nonumber
\end{eqnarray}
For the solution (\ref{ss2}) similar comments to the ones made for (\ref{s3}) can be done. Note that the terms proportional to $Hr^2$ have the same coefficient
for both (\ref{s3}) and (\ref{ss2}). In the general case with  $\ell >2$, we have always the corrections starting from the term $\sim r^2$ 
up to the term $\sim 1/r^{\ell -2}$.  

\section{A preliminary study of the polar perturbations}               

Polar perturbations, as well known, are much more involved with respect to the axial case. We have at our disposal seven equations for the four
variables $N,L,T,V$. In the Minkowski case \cite{13} (and also more generally) only four equations are independent. However, as shown in the sections above,
in the presence of $\Lambda$ we have a mixture of retarded and advanced perturbed coefficients and to have a solution in a closed form, we need of
more than four equations to be managed. In the following we show that this is the case.\\
To start with, we consider the Einstein's equations for the components $(t)(r),(t)(\theta),(r)(\theta)$:
\begin{eqnarray}
& & s\left[L(s,r)-T(s,r)-rT_{,r}(s,r)+(n+1)V(s,r)+r(n+1)V_{,r}(s,r)\right]+\nonumber\\
& & +rHN_{,r}(s,r)=0\label{27}\\
& & s\left[L(s,r)+T(s,r)-V(s,r)\right]=2HN(s,r)\label{28}\\
& & rN_{,r}(s,r)-N(s,r)-L(s,r)+r\left[T_{,r}(s,r)-V_{,r}(s,r)\right]=0.\label{29}
\end{eqnarray}
Equations (\ref{27})-(\ref{29}) do not involve retarded or advanced functions and thus provide three independent equations.
A key equation is the (\ref{28}). The important fact is that (\ref{28}) is an algebraic equation also for
$H\neq 0$. Hence, the (\ref{28}) allows to obtain $T(s,r)$ as a linear algebraic function of 
$L(s,r),V(s,r),N(s,r)$. As a result we have the two independent equations (\ref{27}) and (\ref{29}) in terms of the three 
variables $N(s,r), L(s,r), V(s,r)$. We need of a third independent equation to assure the integrability of the polar equations
in a closed form.\\
For the remaining equations we have: for the component $(\theta)(\theta)-(\phi)(\phi)$:
\begin{eqnarray}
& & (s-2H)V(s-2H,r)\left[s+H\right]=\label{30}\\
& & =V_{,r,r}(s,r)+\frac{2}{r}V_{,r}(s,r)+\frac{L(s,r)+N(s,r)}{r^2},\nonumber
\end{eqnarray}
for the component $(t)(t)$:
\begin{eqnarray}
& & 3r^2H^2 N(s-2H,r)-r^2H(s-2H)L(s-2H,r)-2r^2H(s-2H)T(s-2H,r)+\nonumber\\
& & +2r^2(n+1)H(s-2H)V(s-2H,r)=(2+n)L(s,r)+rL_{,r}(s,r)+nT(s,r)-\nonumber\\
& & -3rT_{,r}(s,r)-r^2T_{,r,r}(s,r)+r(n+1)[rV_{,r,r}(s,r)+3V_{,r}(s,r)],\label{31}
\end{eqnarray}
for the component $(r)(r)$:
\begin{eqnarray}
& & r^2 N(s-2H,r)H\left[s+H\right]-\label{32}\\
& & -r^2(s-2H)T(s-2H,r)\left[s+H\right]
+r^2(n+1)(s-2H)V(s-2H)\left[s+H\right]=\nonumber\\
& & =(n+1)N(s,r)-rN_{,r}(s,r)+L(s,r)+nT(s,r)-\nonumber\\
& & -rT_{,r}(s,r)+r(n+1)V_{,r}(s,r).\nonumber
\end{eqnarray}
Finally, for the component $(\theta)(\theta)+(\phi)(\phi)$ we have:
\begin{eqnarray}
& & 2r^2HN(s-2H,r)\left[s+H\right]-r^2(s-2H)L(s-2H,r)\left[s+H\right]-\label{33}\\
& & -r^2(s-2H)T(s-2H,r)\left[s+H\right]+\nonumber\\
& & +r^2(n+1)(s-2H)V(s-2H,r)\left[s+H\right]=\nonumber\\
& & =2r(n+1)V_{,r}(s,r)+rL_{,r}(s,r)+(n+1)L(s,r)+(n+1)N(s,r)-\nonumber\\
& & -2rT_{,r}(s,r)-r^2N_{,r,r}(s,r)-r^2T_{,r,r}(s,r)-\nonumber\\
& & -rN_{,r}(s,r)+
r^2(n+1)V_{,r,r}(s,r).\nonumber
\end{eqnarray}
Equations (\ref{30})-(\ref{33}) describe 
a complicated system involving advanced perturbed metric coefficients. Nevertheless,  
notice that
in the first members of the (\ref{30})-(\ref{33}), the advanced coefficients $N,L,T,V$ do appear algebraically without 
partial derivatives with respect to $r$.
We can advantage of the algebraic equation (\ref{28}) to obtain $T(s-2H,r)$ as a function of 
$N(s-2H,r),L(s-2H,r),V(s-2H,r)$. The expressions so obtained must be substituted in the (\ref{30})-(\ref{33}): we obtain a linear system of four 
equations in the three unknown functions $N(s-2H,r),L(s-2H,r),V(s-2H,r)$. We can thus use the independent equations
(\ref{30}),(\ref{31}),(\ref{33}) to solve algebraically the functions $N(s-2H,r),L(s-2H,r),V(s-2H,r)$ in terms
of $N(s,r),L(s,r),V(s,r)$ and their partial derivatives with respect to the radial coordinate
$r$. Finally, the expressions so obtained must be inserted in the equation (\ref{32}). Hence we have
three independent equations, (\ref{27}), (\ref{29}) and (\ref{32})\footnote{Note that these are the three independent equations used
	in \cite{8} to integrate the polar equations in the Minkoskian case.}, in terms of non advanced (and retarded) functions 
and solution can be obtained in a closed form.
The expressions involved are quite cumbersome and not illuminating and we leave to a further paper a study of these equations.

\section{Conclusions and final remarks}

Since perhaps we live in a Friedmann flat universe, it is
urgent that we understand the theory of (at least) linearized gravitational waves propagating in an expanding universe. 
To this purpose, we have considered the simple
but physically relevant case of a linearized gravitational wave propagating in an expanding de Sitter universe. Quite recently, in 
\cite{12,13,14,15} the study focused on the effect caused by an expanding universe and on the 
correct asymptotic conditions suitable to identify a gravitational wave. Also in \cite{16} the main purpose has been 
the study of the observable effects
due to $\Lambda$ on the detection of gravitational waves. These studies conclude that the effect, 
thanks to the actual small value for
$\Lambda$, are completely negligible, but in a non-static
non-asymptotically flat background using the analysis performed
in a static asymptotically flat scenario is generally no longer applicable. 
Moreover, in \cite{16} has been shown that the use of the Bondi-Sachs formalism makes it possible to 
write down and solve the perturbations equations in a de Sitter universe in frequencies space. Unfortunately, the de Sitter background is a lucky case since in more general cosmological spacetimes the Bondi-Sachs coordinates are at our disposal only numerically.
Hence, we need of other approaches that are, at least in principle, suitable for a generalization by using usual comoving coordinates or coordinates adatped 
to the sphericity. 
Moreover, in \cite{a,b} the authors derive the axial equations \cite{a} by using the conformal time coordinate instead of the cosmic time
$t$, the time we measure within our cosmoving galaxy, to test the Huygens principle in particular in a dust filled and radiation 
dominated universe. Our purpose is to derive the Regge-Wheeler equation for axial mode in the frequencies domain with respect 
to the cosmological time $t$ and to study the effects on the propagating waves caused by the cosmological 
constant. To this purpose, we shown that 
the use of the Laplace transform with respect to the cosmic time 
$t$ (the time we measure in our galaxy) makes the job. First of all, the Laplace transformed
axial and polar perturbations are functions of retarded and advanced metric coefficients with respect to the 
Laplace parameter $s$.
In the axial case, we remarkable obtain, for the first time, a generalized second
order diferential equation in the usual comoving coordinates and in the frequencies domain as a function of the
advanced Regge-Wheeler function that for $H=0$ reduces to the usual one in a Minkowski
perturbed background.
The fact that the so obtained Regge-Wheeler equation
presents a coupling between $Z(s,r)$ and $Z(s-2H,r))$
implies that, although the observable effects on the detection of a gravitational wave can be negligible for the actual very small cosmological constant, as shown in \cite{12,13,14,15}, non
trivial mathematical and physical issues can arise. 
Moreover, section 5 is an attempt to study the physical meaning of the complex parameter $s$. 
According to our reasonings, a positive abscissa of convergence $s_0=2H=Inf_{\Re(s)}$ 
implies a change of frequency as perceived by a 
comoving observer. In practice, the scale factor $e^{2Ht}$ acts as a translational term in the Laplace language and as 
a consequence a shift term with respect to the frequancy $\omega$ does appear. Stated in other words, the 
expansion factor $e^{2Ht}$ generates a new redshift effect on the frequency $\omega$ given by the (\ref{v3}).\\
Remember that a perturbation Laplace 
transformable with $s_0=0$ is also Fourier transformable and the comparison between Laplace and Fourier modes is 
easily obtained in the limit $s_0=0$. Conversely, a Laplace transformable perturbation with
$s_0>0$ is no longer Fourier transformable and direct comparison it is no longer available.\\  
For $\omega\sim H$, the traveling wave is practically frozen as seen by a comoving observer.
This result can be compared, as explained in section 5, with the usual results present in the literature concerning the modes 
during primordial inflation in terms of the proper wavenumber ${\bf k}/a(t)$. In fact, note that the quanta composing a 
gravitational wave are massless gravitons with wavenumber given by $|{\bf k}|\sim 2\pi/{{\lambda}_c}$ with 
${\lambda}_c$ the comoving wavelength related to the proper wavelength ${\lambda}_p$ by the usual relation
${\lambda}_p=a(t){\lambda}_c$. For massless gravitons we thus have, thanks to the (\ref{v3}):
\begin{equation}
{\omega}_p^2=\omega^2-\frac{4}{3}\Lambda=\frac{4\pi^2}{{\lambda}_p^2}.
\label{c1}
\end{equation} 
From the (\ref{c1}), it follows that, for ${\lambda}_p>>1/H$ (super Hubble horizon modes) we have ${\omega}\sim 2H$ and 
${\omega}_p=o(H)=o(\sqrt{\Lambda})$. For the actual very low value for $\Lambda$ for these modes we practically 
have ${\omega}_p\simeq 0$, i.e. the modes are practically frozen and near a Bose-Einstein condensate phase.
Conversely, modes with ${\lambda}_p<<1/H$ have ${\omega}>>H$ and as a result ${\omega}_p\simeq\omega$, i.e. modes well within
the Hubble horizon practically do not feel the exponential expansion of the universe. All these facts, as indicated in section
5, are in agreement with well known results in a de Sitter universe. An advantage of our approach is to obtain a shift 
effect on the frequencies space as measured with respect to the proper comoving time $t$.

This fact open the possibility, as discussed in section 5, that the cosmological constant is made of gravitons at a mean frequency
$\overline{\omega}$ of the order of $H$ \cite{vs}. 
 
Also the polar perturbation equations are studied. We have shown that the system of seven equations
allows, after simple but lenghly calculations, to express the relevant equations in terms of
$N(s,r),L(s,r),V(s,r)$.
 
Our formalism also open the possibility to study the problem of quasi-normal modes of a generic astrophysical source emebedded 
in a Friedmann universe in a similar way to the approach present in \cite{3}-\cite{10}.
The complex parameter $s$ can be obtained by solving a suitable eigenvalue problem. The real part of
$s$ represents the frequency of quasi-normal modes with the $Inf_{\Re(s)}$ representing the eventual shift 
contribution for the frequency perceived
by distant comoving observers. 

It should also be mentioned the fact that our formalism might allow the study of fluctuations at the primordial inflation in the frequencies space. 

\section*{Acknowledgements}  
I would like to thank Tommaso Isola and Giuseppe Ruzzi for useful discussions and suggestions.

\end{document}